\documentclass[]{raa}            
\usepackage{graphicx,times}
\usepackage{natbib}
\usepackage{multirow}

\providecommand{\bm}[1]{\mbox{\boldmath$#1$\unboldmath}}

\begin{document}
\title{Influence of Coronal Holes on CMEs in Causing SEP Events}

\author{ Chenglong Shen \inst{1,2}\and Jia Yao\inst{1} \and Yuming Wang\inst{1}\and Pinzhong Ye\inst{1},
 X. P. Zhao\inst{3} \and S. Wang\inst{1}}
 \institute{ CAS Key Laboratory of Basic Plasma Physics, School of Earth
\& Space Sciences, University of Science \& Technology of China,
Hefei, Anhui 230026, People's Republic of China
 (clshen@ustc.edu.cn, ymwang@ustc.edu.cn)
 \and State Key Laboratory of Space Weather, Chinese Academy of Science,
Beijing, 100080, China.
 \and W. W. Hansen Experimental Physics Laboratory, Stanford University, Stanford,
CA 94305.}

\abstract{
The issue of the influence of coronal holes (CHs) on coronal mass
ejections (CMEs) in causing solar energetic particle (SEP) events is
revisited. It is a continuation and extension of our previous work
\citep{Shen_etal_2006}, in which no evident effect of CHs on CMEs in
generating SEPs were found by statistically investigating 56 CME
events. This result is consistent with the conclusion obtained by
Kahler in 2004. In this paper, we extrapolate the coronal magnetic field,
define CHs as the regions consisting of only open magnetic field
lines and perform a similar analysis on this issue for totally 76 events by extending
the study interval to the end of 2008. Three key
parameters, CH proximity, CH area and CH relative position, are
involved in the analysis. The new result confirms the previous conclusion
that CHs did not show any evident
effect on CMEs in causing SEP events.
\keywords{acceleration of particles --- Sun: coronal mass ejections
--- Sun: coronal holes --- Sun: particle emission}}
 \authorrunning{C. L. Shen, J. Yao, Y. M. Wang, P. Z. Ye, X. P. Zhao and S. Wang}
   \titlerunning{CH influence SEP events }  
   \maketitle

\section{Introduction}
Gradual solar energetic particle (SEP) events are thought to be a
consequence of CME-driven shocks generating plenty of SEPs which
would be observed near the Earth. In our previous work in 2006, we
statistically studied the effect of coronal holes (CHs) on the CMEs
causing SEP events by investigating the CME source locations and
their relation with the CHs identified in EUV 284 \AA$ $\citep[][
hereafter Paper I]{Shen_etal_2006}. It was implied that neither CH
proximity nor CH relative location exhibits any evident effect on
the intensities of SEP events. This result is consistent with the
conclusion obtained by \citet{Kahler_2004}, who comparatively
studied the SEP events produced in the fast and slow solar wind streams
and found no significant bias against SEP
production in fast-wind regions which are believed to originate from
CHs.

These findings seem not quite fit people's `common sense',
because CHs are believed to be regions with low-density and low
temperature in the corona \citep[e.g.][]{Harvey_Recely_2002}, from
which the solar wind is fast and the magnetic field is open, and
therefor apparently three disadvantages for a CME to produce SEP
may exist when it is near a coronal hole region. These advantage
are: (1) the background solar wind speed $V_{sw}$ near CHs is larger
than that in other regions; (2) the plasma density near CHs is much
lower than that in other regions, so that the Alfv\'{e}n speed $V_a$
is larger \citep{Shen_etal_2007,Gopalswamy_etal_2008}; and (3) the
magnetic field lines in CHs are open. The first two disadvantages
suggest that a strong shock might be hardly produced near CHs. The
third one implies that particles might be able to escape from the
shock acceleration process earlier and easier. Thus, it can be
expected that CHs would influence the CME in producing SEP events.
The work by \citet{Kunches_Zwickl_1999} was consistent with the
picture depicted above. In their paper,
they found that the CH may
delay the onset times of SEPs when a CH is present between
Sun-observer line and the solar source of the SEP event.
They also speculate that the peak intensity could be influenced by CH.
However, they did not statistical study such influence.
It is hard to say that their conclusion is statistically significant.

In principle, CHs are open field regions, though they were first
identified in observations \citep[e.g.][]{Zirker_1977}.
\citet{Kunches_Zwickl_1999} identified CHs based on He 10830 \AA$ $.
In our 2006 work (Paper I), CHs were auto-determined based on  EUV
284\AA$ $ images taken by SOHO/EIT. Thus, it is doubtable whether or
not the CHs identified in EUV wavelengths really represent open
field regions. Another doubt in our 2006 work is that only frontside
CHs are taken into account.
In order to remove the doubt and get a more reliable
result, we look into this topic again by extrapolating coronal
magnetic field instead of analyzing EUV images. The term `CHs' in
this paper therefore actually refers to open field regions. The
magnetic field extrapolation and determination of CHs are introduced
in Section 2. Section 3 presents the statistical analysis. A brief
summary and conclusions are given in Section 4.

\section{Determination of coronal holes}
So far, there are no observations of coronal magnetic field. Most
information of coronal magnetic field comes from various
extrapolation techniques \citep[e.g.][]{Schatten_etal_1969,Altschuler_Newkirk_1969,Schatten_1971,Zhao_Hoeksema_1992,Zhao_Hoeksema_1994,Zhao_Hoeksema_1995,Zhao_etal_2002a}.
In this paper, the current sheet-surface source (CSSS) model developed by
Zhao and his colleagues \citep{Zhao_Hoeksema_1995,Zhao_etal_2002a}
will be used to extrapolate the coronal magnetic field and identify
the coronal hole regions. In our calculation, the daily-updated
synoptic charts of photospheric magnetic field from the Michelson
Doppler Imager (MDI \citep{Scherrer_etal_1995}) on board SOHO
spacecraft is adopted as the bottom boundary condition, the
extrapolated global magnetic field is a kind of average over the
carrington rotation, and may not exactly reflect the state at the
time of interest. However, because CHs are long-lived structures in
the solar atmosphere, we think that such approximation of global
field would not significantly distort our results. To determine
where are open field regions, we design 180-by-90 grid points (a
point every 2 degree in longitude and 1/45 in sine latitude) at
photosphere as the roots of magnetic field lines. In other words, a
total of 16200 field lines will be traced to check if they are open
or closed.

By using this method, CHs are defined as the regions consisting of
open magnetic field lines on the photosphere. Neighboring regions
with a spherical separate distance $\le7.5^\circ$ are grouped as one
region. Those small regions with area less than 0.0024 $A_s$  were
discarded to raise the credibility of the determined CHs. Here $A_s$
is the total area of solar surface. The size of 0.0024 $A_s$ is
about a $10^\circ\times10^\circ$ grid at the center of the solar
disk (the projection of the Sun on the plane of sky). The projection
effect has been corrected in the calculation of the area of open
magnetic field regions. Compared with the previous approach
developed by \citet{Shen_etal_2006}, this method can not only
obtain all CHs over the full solar surface (not just those on the front-side solar disk),
but also dig out the CHs covered by some bright structures (e.g., active regions)
in EIT 284 \AA$ $ images.
\begin{figure*}[tb]
\begin{center}
\includegraphics[width=\hsize]{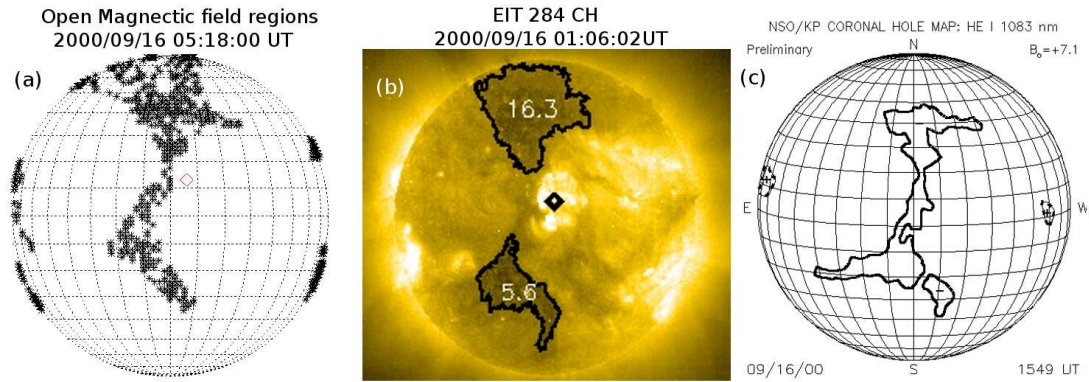}
\caption{(a) An example on 2000 September 16 showing CH
determination by our method,
the regions marked by crosses are the determined CHs, and the
diamond indicates the CME location.
(b) The corresponding EIT 284\AA$ $
image superimposed with CH boundaries obtained by the method in
Paper I, (c) Kitt Peak CH map.} \label{ch}
\end{center}
\end{figure*}

Figure \ref{ch} shows an example on 2000 September 16, which was
also presented in Paper I. The  asterisks in Figure \ref{ch}(a) denote
the open field regions inferred by the method (the carrington map
has been re-mapped on the solar sphere). Figure \ref{ch}(b) and Figure \ref{ch}(c) show
the corresponding EIT 284\AA$ $ image over-plotted with the CH
boundaries determined in Paper I and Kitt Peak CH
map for comparison.

It is obvious that CHs obtained here
are similar to, but not the same, as those in the other two.
The CHs presented in the EIT 284\AA$ $ is in high corona and the Kitt
Peak CHs is in lower corona\citep{Harvey_Recely_2002} ,
whereas our extrapolated CHs is on the photosphere.
As CHs may expand rapidly and superradially with increasing height
\citep{Munro_Jackson_1977,Fisher_Guhathakurta_1995,DeForest_etal_2001},
the difference in altitude between them probably is one of the major
causes of the apparent difference in the CH shape. The regions
determined here could be treated as the roots of CHs. Besides, the CHs
at east and west limbs in Figure \ref{ch}(a) and Figure \ref{ch}(c) can not be recognized in
 Figure \ref{ch}(b). This is because of the shielding of the brightness of nearby active region.
In addition, the same CH also exhibit different shapes and properties at different panels.
The big CH extended from north to south in center longitude region shown in Figure \ref{ch}(a) and Figure \ref{ch}(c) has been divided into two separated CHs in Figure \ref{ch}(b).
This may also because that the brightness of the active region shield the dark region located at solar center, which makes this big CH seems like two isolated dark region.

\section{Statistical results}
In this paper, the time period of 1997 -- 2003 we used in paper I is extended to the end of 2008.
All fast halo CME events originating from west hemisphere
during this period are studied.
As the same as we did in Paper I, the `fast' and `halo' mean that the CME projected
speed measured in SOHO/LASCO is larger than 1000 km/s and the span angle is larger
than $130^\circ$. Since the daily-updated magnetic field synoptic chart
on 1998 November 5 is not available for use, the event occurred on that day is excluded. Thus
a total of 76 events will be analyzed. Table 1 lists the events
including the parameters of CMEs, CHs and SEPs. The key parameters
we used to analyze the effect of CHs on CMEs in producing SEPs are
the CH proximity (column 6), the area of the CH nearest to the CME
(column 7) and the relative position of the CH (column 8).
All parameter have the same meaning as those in Paper I.

\begin{table}[tb]
\centering \caption{Frontside fast halo CMEs originating from the
west hemisphere during 1997 -- 2008}
\label{tb_statistics}
\tiny
\begin{tabular}{c|cccc|ccc|cc}
\hline {No.} &\multicolumn{4}{|c|}{CME$^a$}
&\multicolumn{3}{|c|}{CH}
&\multicolumn{2}{|c}{SEP}\\
 &Date Time &Width &Speed &Location$^b$ & $Proximity^c$& $Area^d$&$P^e$
 &$\ge$10MeV$^f$&$\ge$50MeV$^g$\\
\hline
 1&1997-11-06, 12:10:41&   360&1556&S18,W62&0.60(D)&0.0049(a)&N&  490.0& 116.0\\
 2&1998-04-20, 10:07:11&   165&1863&S47,W70&0.28(d)&0.0041(a)&Y& 1610.0& 103.0\\
 3&1998-05-06, 08:29:13&   190&1099&S15,W68&0.54(D)&0.0066(A)&Y&  239.0&  19.3\\
 4&1999-06-04, 07:26:54&   150&2230&N19,W85&0.86(D)&0.0053(a)&Y&   64.0&   0.9\\
 5&1999-06-28, 21:30:00&   360&1083&N23,W42&0.51(D)&0.0070(A)&Y&   -1.0&  -1.0\\
 6&1999-09-16, 16:54:00&   147&1021&N42,W30&0.31(d)&0.0045(a)&Y&   -1.0&  -1.0\\
 7&2000-02-12, 04:31:00&   360&1107&N13,W28&0.51(D)&0.0116(A)&Y&    2.7&  -1.0\\
 8&2000-04-04, 16:32:00&   360&1188&N16,W60&0.07(d)&0.0034(a)&N&   55.8&   0.3\\
 9&2000-05-15, 16:26:00&$>$165&1212&S23,W68&0.87(D)&0.0028(a)&Y&    1.0&  -1.0\\
10&2000-06-10, 17:08:00&   360&1108&N22,W40&0.25(d)&0.0059(a)&Y&   46.0&   6.5\\
11&2000-06-25, 07:54:00&   165&1617&N10,W60&0.31(d)&0.0066(A)&N&    4.6&  -1.0\\
12&2000-06-28, 19:31:00&$>$134&1198&N24,W85&0.05(d)&0.0119(A)&Y&   -1.0&  -1.0\\
13&2000-07-14, 10:54:00&   360&1674&N17,W 2&0.97(D)&0.0101(A)&Y&24000.0&1670.0\\
14&2000-09-12, 11:54:00&   360&1550&S14,W 6&0.69(D)&0.0037(a)&Y&  321.0&   2.0\\
15&2000-09-16, 05:18:00&   360&1215&N13,W 6&0.11(d)&0.0296(A)&Y&    7.1&  -1.0\\
16&2000-11-08, 23:06:00&$>$170&1738&N14,W64&0.88(D)&0.0045(a)&N&14800.0&1880.0\\
17&2000-11-24, 15:30:00&   360&1245&N21,W12&0.30(d)&0.0025(a)&Y&   94.0&   5.0\\
18&2001-02-11, 01:31:00&   360&1183&N21,W60&0.19(d)&0.0056(a)&N&   -1.0&  -1.0\\
19&2001-04-02, 22:06:00&   244&2505&N16,W65&0.54(D)&0.0103(A)&Y& 1110.0&  53.5\\
20&2001-04-09, 15:54:00&   360&1192&S20,W 4&1.06(D)&0.0081(A)&Y&    5.9&   1.2\\
21&2001-04-10, 05:30:00&   360&2411&S20,W10&1.07(D)&0.0065(A)&Y&  355.0&   3.7\\
22&2001-04-12, 10:31:00&   360&1184&S20,W43&1.00(D)&0.0102(A)&Y&   50.5&   5.8\\
23&2001-04-15, 14:06:00&   167&1199&S20,W85&1.03(D)&0.0111(A)&N&  951.0& 275.0\\
24&2001-04-26, 12:30:00&   360&1006&N23,W 2&0.83(D)&0.0128(A)&Y&   57.5&  -1.0\\
25&2001-07-19, 10:30:00&   166&1668&S 9,W61&0.36(D)&0.0033(a)&Y&   -1.0&  -1.0\\
26&2001-10-01, 05:30:00&   360&1405&S20,W89&0.25(d)&0.0054(a)&Y& 2360.0&  24.5\\
27&2001-10-22, 15:06:00&   360&1336&S18,W20&1.02(D)&0.0081(A)&N&   24.2&   2.5\\
28&2001-10-25, 15:26:00&   360&1092&S18,W20&0.32(D)&0.0049(a)&Y&   -1.0&  -1.0\\
29&2001-11-04, 16:20:00&   360&1274&N 6,W18&0.58(D)&0.0036(a)&Y&31700.0&2120.0\\
30&2001-11-22, 23:30:00&   360&1437&S17,W35&0.14(d)&0.0046(a)&N&18900.0& 162.0\\
31&2001-12-26, 05:30:00&$>$212&1446&N 9,W61&0.27(d)&0.0047(a)&N&  780.0& 180.0\\
32&2002-04-17, 08:26:00&   360&1218&N13,W12&0.22(d)&0.0068(A)&Y&   24.1&   0.4\\
33&2002-04-21, 01:27:00&   241&2409&S18,W79&0.06(d)&0.0081(A)&Y& 2520.0& 208.0\\
34&2002-05-22, 03:50:00&   360&1494&S15,W70&0.73(D)&0.0065(A)&Y&  820.0&   1.1\\
35&2002-07-15, 20:30:00&   360&1132&N20,W 2&0.08(d)&0.0164(A)&Y&  234.0&   0.9\\
36&2002-07-18, 08:06:00&   360&1099&N20,W33&0.05(d)&0.0098(A)&Y&   14.2&   0.6\\
37&2002-08-06, 18:25:00&   134&1098&S38,W18&0.31(d)&0.0040(a)&Y&   -1.0&  -1.0\\
38&2002-08-14, 02:30:00&   133&1309&N10,W60&0.04(d)&0.0083(A)&N&   26.4&  -1.0\\
39&2002-08-22, 02:06:00&   360&1005&S14,W60&0.46(D)&0.0035(a)&N&   36.4&   6.0\\
40&2002-08-24, 01:27:00&   360&1878&S 5,W89&0.28(d)&0.0041(a)&Y&  317.0&  76.2\\
41&2002-11-09, 13:31:00&   360&1838&S 9,W30&0.42(D)&0.0105(A)&Y&  404.0&   1.5\\
42&2002-12-19, 22:06:00&   360&1092&N16,W10&0.58(D)&0.0170(A)&Y&    4.2&  -1.0\\
43&2002-12-21, 02:30:00&   225&1072&N30,W 0&0.75(D)&0.0190(A)&Y&   -1.0&  -1.0\\
44&2002-12-22, 03:30:00&   272&1071&N24,W43&0.69(D)&0.0224(A)&Y&   -1.0&  -1.0\\
45&2003-03-18, 12:30:00&   209&1601&S13,W48&0.14(d)&0.0199(A)&Y&    0.8&  -1.0\\
46&2003-03-19, 02:30:00&   360&1342&S13,W56&0.17(d)&0.0212(A)&N&   -1.0&  -1.0\\
47&2003-05-28, 00:50:00&   360&1366&S 5,W25&0.12(d)&0.0044(a)&Y&  121.0&   0.3\\
48&2003-05-31, 02:30:00&   360&1835&S 5,W65&0.18(d)&0.0034(a)&Y&   27.0&   2.3\\
49&2003-10-26, 17:54:00&$>$171&1537&N 3,W43&0.15(d)&0.0032(a)&Y&  466.0&  10.4\\
50&2003-10-27, 08:30:00&$>$215&1380&N 3,W48&0.08(d)&0.0028(a)&Y&   52.0&   9.6\\
51&2003-10-29, 20:54:00&   360&2029&S16,W 5&0.72(D)&0.0027(a)&Y& 2470.0& 389.0\\
52&2003-11-02, 09:30:00&   360&2036&S16,W51&0.07(d)&0.0035(a)&N&   30.0&   0.8\\
53&2003-11-02, 17:30:00&   360&2598&S16,W56&0.08(d)&0.0035(a)&N& 1570.0& 155.0\\
54&2003-11-04, 19:54:00&   360&2657&S16,W83&0.07(d)&0.0037(a)&Y&  353.0&  15.3\\
55&2003-11-11, 13:54:00&   360&1315&S 3,W63&0.24(d)&0.0049(a)&Y&   -1.0&  -1.0\\
56&2004-04-08, 10:30:19&   360&1068&S16,W 6&0.02(d)&0.0033(a)&Y&   -1.0&  -1.0\\
57&2004-07-25, 14:54:05&   360&1333&N 3,W33&0.72(D)&0.0028(a)&Y&   54.6&   0.8\\
58&2004-07-29, 12:06:05&   360&1180&N 0,W89&0.32(D)&0.0060(a)&Y&   -1.0&  -1.0\\
59&2004-07-31, 05:54:05&   197&1192&N 9,W89&0.37(D)&0.0059(a)&Y&   -1.0&  -1.0\\
60&2004-11-07, 16:54:05&   360&1759&N 9,W16&0.38(D)&0.0220(A)&Y&  495.0&   4.7\\
61&2004-11-09, 17:26:06&   360&2000&N 8,W48&0.31(d)&0.0270(A)&Y&   82.4&   0.9\\
62&2004-11-10, 02:26:05&   360&3387&N 7,W53&0.12(d)&0.0259(A)&Y&  424.0&  13.5\\
63&2004-12-03, 00:26:05&   360&1216&N 9,W 1&0.35(D)&0.0028(a)&Y&    3.2&  -1.0\\
64&2005-01-15, 23:06:50&   360&2861&N13,W 3&0.76(D)&0.0112(A)&Y&  365.0&  12.8\\
65&2005-01-17, 09:30:05&   360&2094&N13,W20&0.46(D)&0.0245(A)&Y&  269.0&   4.0\\
66&2005-01-17, 09:54:05&   360&2547&N13,W20&0.46(D)&0.0245(A)&Y& 5040.0& 387.0\\
67&2005-01-19, 08:29:39&   360&2020&N13,W45&0.51(D)&0.0246(A)&Y&   -1.0&  -1.0\\
68&2005-02-17, 00:06:05&   360&1135&S 1,W19&0.02(d)&0.0059(a)&Y&   -1.0&  -1.0\\
69&2005-07-09, 22:30:05&   360&1540&N 9,W29&0.31(d)&0.0313(A)&Y&    3.0&  -1.0\\
70&2005-07-13, 14:30:05&   360&1423&N 9,W76&0.33(D)&0.0382(A)&Y&   12.5&   0.3\\
71&2005-07-14, 10:54:05&   360&2115&N 9,W87&0.39(D)&0.0395(A)&Y&  134.0&   2.6\\
72&2005-08-22, 01:31:48&   360&1194&S12,W51&0.18(d)&0.0027(a)&Y&    7.3&  -1.0\\
73&2005-08-22, 17:30:05&   360&2378&S12,W60&0.18(d)&0.0027(a)&N&  337.0&   4.8\\
74&2005-08-23, 14:54:05&   360&1929&S13,W75&0.24(d)&0.0035(a)&Y&   -1.0&  -1.0\\
75&2006-12-13, 02:54:04&   360&1774&S 8,W19&0.14(d)&0.0027(a)&Y&  698.0& 239.0\\
76&2006-12-14, 22:30:04&   360&1042&S10,W42&0.19(d)&0.0032(a)&Y&  215.0&  13.5\\
\hline\end{tabular}\\
$^a$ Obtained from CME CATALOG (http://cdaw.gsfc.nasa.gov/CME\_list/). \\
$^b$ CME locations determined by the EIT movie. \\
$^c$ Shortest surface distance between a CME and a CH (from the CME
site to the
CH boundary) in units of $R_\odot$, called CH-proximity. `D' means
CH proximity larger than 0.3 $R_s$ while `d' means others.\\
$^d$ Area of the closest CH in units of $A_s$, the area of solar
surface. `A'
means the CH area larger than 0.0061 $A_s$ while `a' means it smaller than 0.0061 $A_s$.\\
$^e$ Relative location of a CH to the corresponding CME. `Y' means
the CH extending into the longitudes between the CME and the field
lines connecting Earth to the Sun
at about $W60^\circ$, and `N' indicates the CH outside the two longitudes. \\
$^f$ Peak fluxes of $\geq10$ MeV-protons in units of pfu, dots mean no SEP observed.\\
$^g$ Peak fluxes of $\geq50$ MeV-protons in units of pfu.
\end{table}

It should be noted that the parameters of CHs we obtained in this paper
were differ from Paper I, which may be caused by the following reasons:
\begin{enumerate}
	\item The nearest CHs for large number of events were changed:
    \begin{enumerate}
    \item As shown in Figure \ref{ch}, the dark regions of CHs shielded by the brightness of active region in
	EIT 284\AA$ $ images can be obtained in this paper.
    This makes the nearest CHs change in 26 events.
    \item  CHs located in solar limb and backside has also been taken into account in this paper as we discussed in Section 2.
      In this paper, the nearest CHs changed to the limb or backsied CHs in totally 14 events.
    \end{enumerate}
	\item For other 15 events, same CHs in this paper and paper I were used.
It is found that tha areas of these 15 CHs were smaller than we obtained in paper I.
In this paper, the CH we obtained can be treated
as the roots of CHs.  As  CHs may expand rapidly and superradially with increasing height
	\citep{Munro_Jackson_1977,Fisher_Guhathakurta_1995,DeForest_etal_2001}, such result could be expected.
	\end{enumerate}
Such variations make the properties of nearest CHs changed largely.
As we discussed before, the nearest CHs in totally 40 events changed. 
Even for same CH, the difference CH shape and different CH height also makes the properties of near CHs change.
In this paper, the relative position of 26 events changed, and 20 events in which were changed from `N' to  `Y'. 
Because of variation of the nearest CHs and the height and shape of same CHs, the group of CH area and CH proximity would hard to be compared.

For simplicity and reliability, we binarize the key parameters
before further analysis. The events with CH proximity larger than
0.31 $R_s$ are marked as `D' and the others marked as `d'. The events
with the CH area larger/smaller than 0.0061 $A_s$ are marked as
`A'/`a'. The parameter of the CH relative position is already
bi-valued.  The separation values 0.31 $R_s$ and 0.0061 $A_s$ are
chosen to make the events near-equally divided into two groups for
the CH proximity and area, respectively. In the following
subsections, we will present the analysis on these difference parameters.

\subsection{The dependence of CH proximity}
\begin{figure*}[tb]
\begin{center}
\includegraphics[width=0.8\hsize]{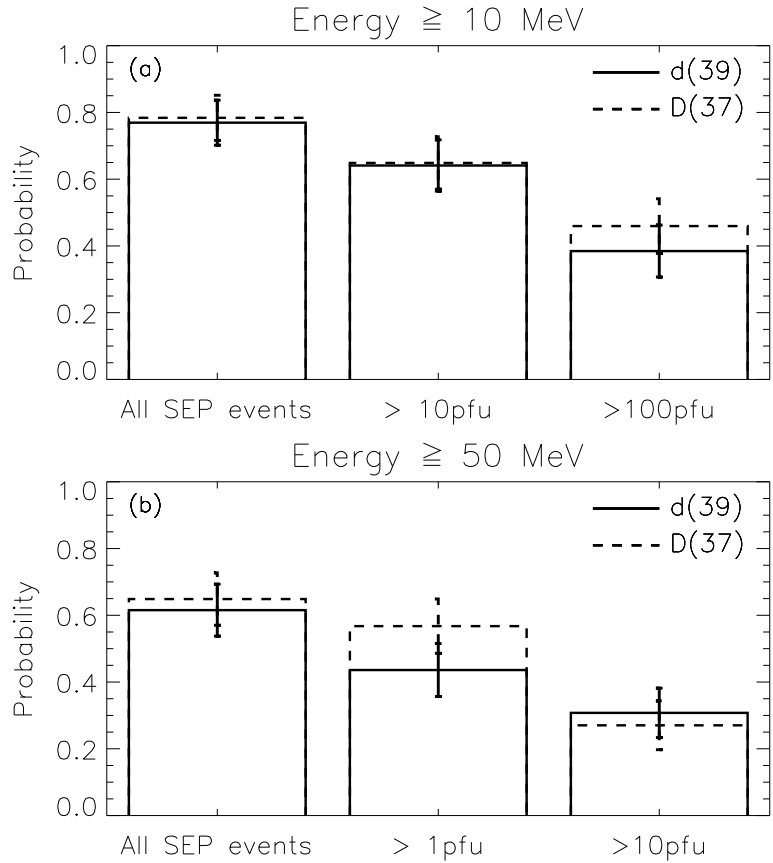}
\caption{Occurrence probabilities, $P$, of SEP events in terms of
the CH proximity for proton energies $\ge$10 MeV and $\ge$50 MeV,
respectively. The probabilities at different groups are indicated by
solid and dashed lines with error bars, respectively.
Difference bins show the probabilities at different flux levels. For
the SEP at energies $\ge$10 MeV, three levels are all SEP events,
$\ge$10 and$\ge$100 pfu events, in which 1 pfu = 1 particle
cm$^{-2}$ s$^{-1}$ sr$^{-1}$. For the SEP at energies $\ge$50 MeV,
they are all SEP events, $\ge$1 and$\ge$10 pfu events. } \label{d}
\end{center}
\end{figure*}

Figure \ref{d} shows the occurrence probabilities, $P$, of SEP
events in terms of the CH proximity for proton energies $\ge$10 MeV
(Panel a) and $\ge$50 MeV (Panel b). The SEP events at difference
flux levels are presented by difference bins. For the SEP event with
proton energy $\ge$ 10 MeV, the three levels are all SEP events, SEP
events with proton flux  $\ge$ 10 pfu and $\ge$ 100 pfu, in which 1
pfu = 1 particle cm$^{-2}$ s$^{-1}$ sr$^{-1}$. For the SEP event
with proton energy $\ge$ 50 MeV, they are all SEP events, SEP events
with proton flux $\ge$ 1 pfu and $\ge$ 10 pfu. Different lines
show the probabilities at different groups. The
probabilities at group `d' and `D' are indicated by solid and
dashed lines with error bars, respectively. The CME number in
each group is marked in the bracket at the top right of the figure.
The error bars indicate the one standard
deviation ($\sigma$) level, which is given by $ \sigma=\sqrt{P(1-
P)/N}$, where N is the total number of CME events for the
corresponding bin.

It is found that the difference of occurrence probabilities of SEP events
between these two groups are small for all flux and energy levels.
All differences between these two groups are less than the value of on standard
deviation (1$\sigma$). Such analysis confirm the result we obtained in paper I that
CHs proximity have no evident effect on CMEs in producing SEP events.

\begin{figure*}[tb]
\begin{center}
\includegraphics[width=0.8\hsize]{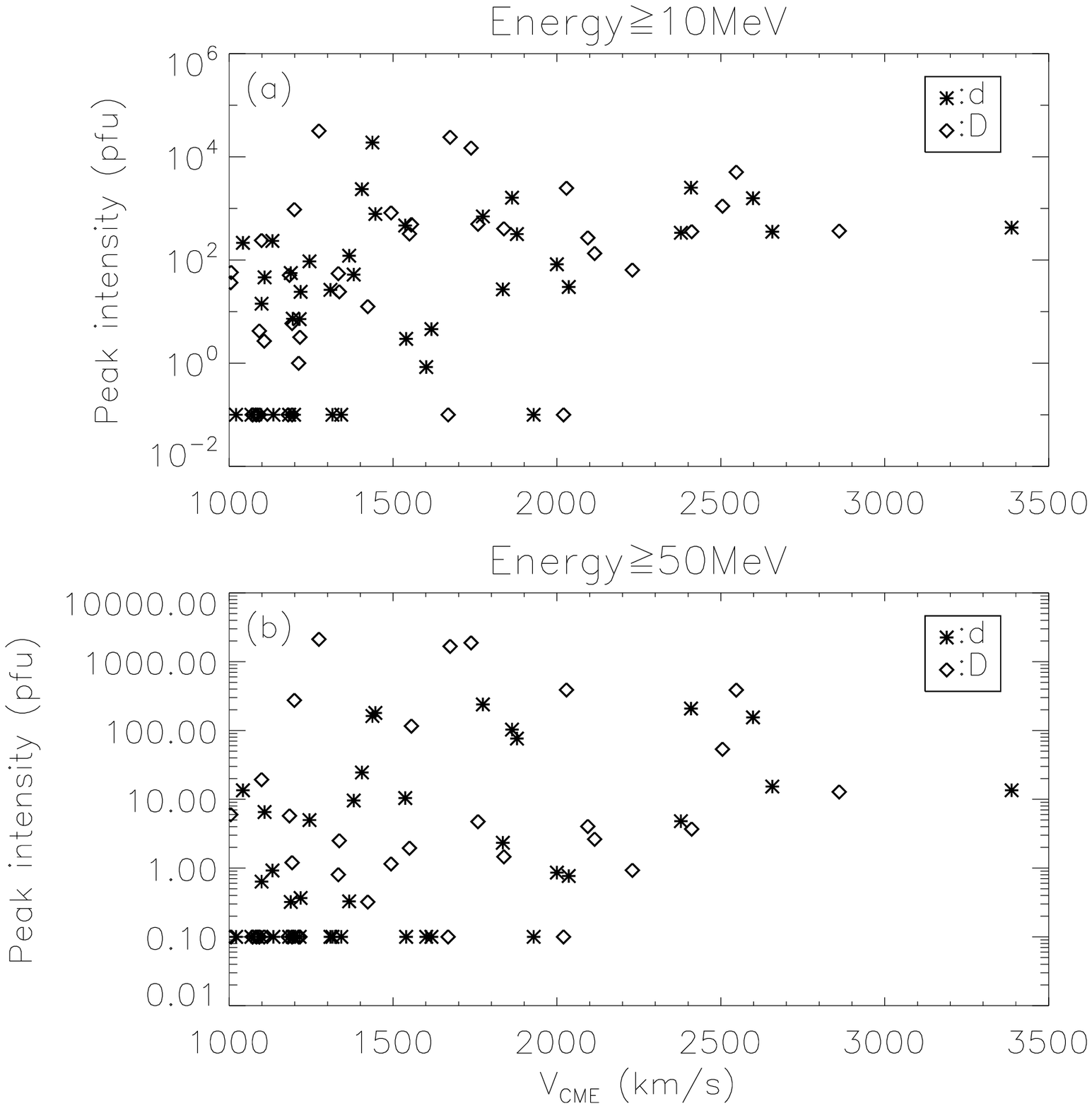}
\caption{The peak intensity of proton with energy $\ge$ 10MeV vs. associated CME speed
for proton energy $\ge$10 MeV (Panel a) and $\ge$ 50 MeV (Panel b).
The asterisks show the CME events in group `d' while diamonds show the CME events
 in group `D'. Points at peak intensity of 0.01 means no SEP associated.
 } \label{d_v}
\end{center}
\end{figure*}
Further, the correlation between the peak intensities of SEP events
and the speed of associated CMEs is studied
(shown in Figure \ref{d_v}). Asterisks in Figure \ref{d_v} show the events in group
`d' and diamonds show the events in group `D'. Points at peak intensity
of 0.01 means no SEP event associated (called as SEPNCMEs in short).
Panel (a) and Panel (b) in this figure shows the events with proton energy $\ge$10MeV and
$\ge$50MeV, respectively.
From this figure, it is found that
the SEP associated CMEs (called as SEPYCMEs in short) were faster than SEPNCMEs.
Almost all (15/16) extremely fast CMEs with speed $\ge2000km/s$ were associated
with SEP events.

\begin{table}[tb]
\centering \caption{ Mean value of CME speed for different groups (in unit of $km.s^{-1}$)}
\label{tb_cmev}
\begin{tabular}{c|c|cc|cc|cc}
\hline
\multirow{2}*{Energy}&\multirow{2}*{SEP}&\multicolumn{2}{|c|}{CH Proximity}&
\multicolumn{2}{|c|}{CH area}&\multicolumn{2}{|c}{Relative Position}\\
\cline{3-8}
&&d&D&a&A&Y&N\\
\hline
\multirow{2}*{$\ge$10 MeV}&Y&1663$\pm$560$^a$&1623$\pm$524&1604$\pm$454&1682$\pm$614
&1655$\pm$559&1603$\pm$474\\
\cline{2-8}
&N&1254$\pm$274&1297$\pm$353&1262$\pm$282&1298$\pm$369&1276$\pm$324&1263$\pm$112\\
\hline
\multirow{2}*{$\ge$50 MeV}&Y&1726$\pm$605&1727$\pm$518&1650$\pm$459&1817$\pm$655
&1755$\pm$574&1629$\pm$511\\
\cline{2-8}
&N&1318$\pm 251$&1232$\pm$288&1250$\pm$249&1305$\pm$292&1264$\pm$280&1363$\pm$183\\
\hline
\end{tabular}\\
$^a$ The number after $\pm$ shows the standard variation.
\end{table}

Table \ref{tb_cmev} gives the comparison of the speed of CMEs at different groups.
Difference columns show the mean value of CME speed at different groups binarized by CH proximity,
CH area and relative position respectively.
The first and second rows show the value of SEPYCMEs  and SEPNCMEs
for the SEP event with proton energy $\ge$10 MeV ,
while the third and forth rows show them for proton energy $\ge$ 50 MeV respectively.

Third and forth columns of Table \ref{tb_cmev} shows the comparison of CME speed at
different groups binarized by CH proximity (group `d' and `D').
It is found that the speed of SEPYCMEs in group `d' and `D' are almost the same.
Meanwhile, the speed of SEPNCMEs in these two groups are also similar.
Such results imply that no significant fast CMEs were required
for producing SEP events when CMEs close to CHs. This result is consistent with
 \citet{Kahler_2004}'s result that no significant fast CME were required for
 producing the SEP events in fast solar wind region.

\subsection{The dependence of CH area}
\begin{figure*}[tb]
\begin{center}
\includegraphics[width=0.8\hsize]{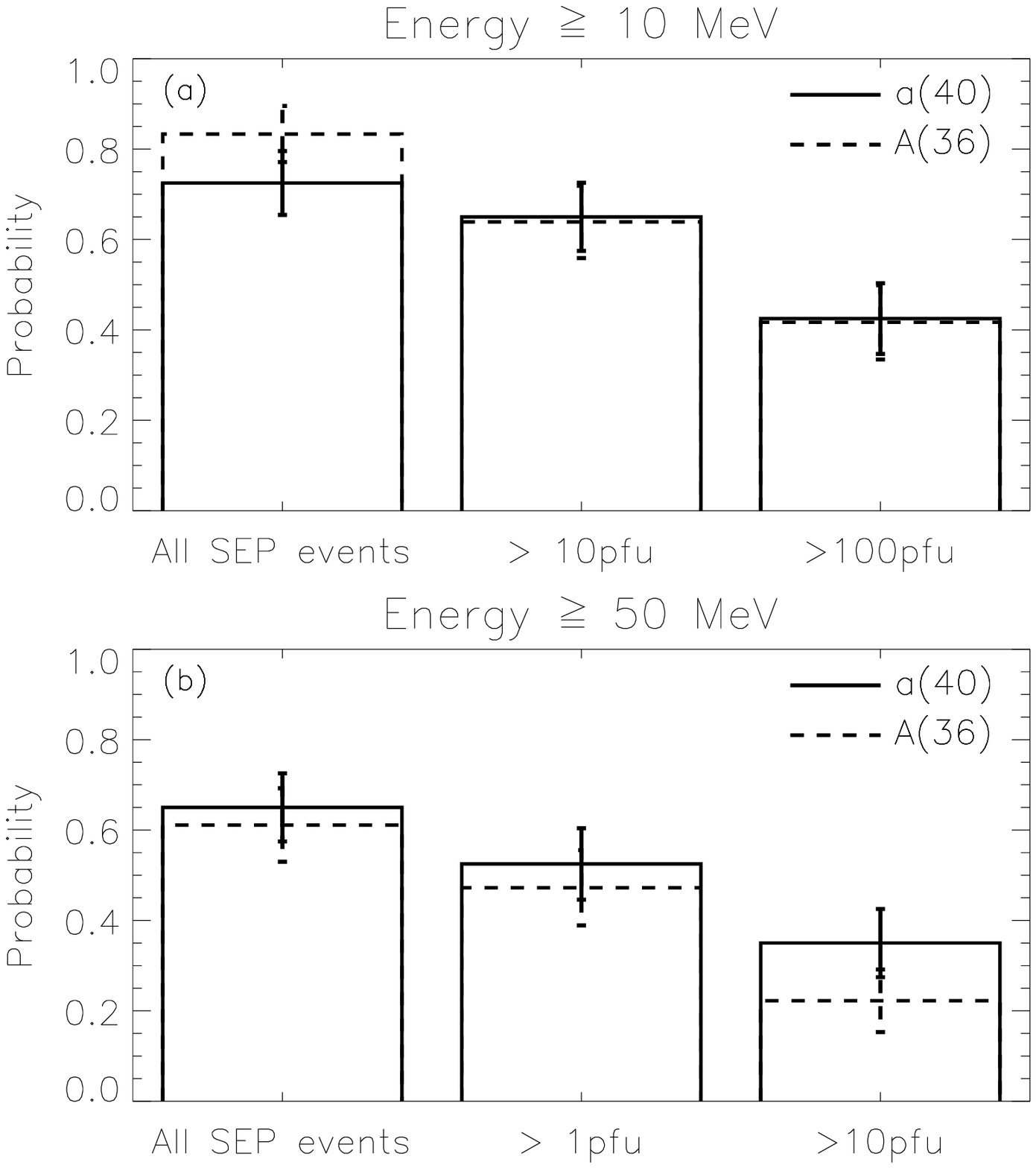}
\caption{Occurrence probabilities, $P$, of SEP events in terms of CH
area for proton energies $\ge$10 MeV and $\ge$50 MeV, respectively.}
\label{a}
\end{center}
\end{figure*}

Figure \ref{a} shows the occurrence probabilities, $P$, of SEP events
in terms of the closest-CH area for proton energies $\ge$10 MeV and
$\ge$50 MeV. For the SEP events with proton energies $\ge$ 10 MeV shown as Figure \ref{a}(a),
the occurrence probabilities of SEP events in group `A' are smaller than them in group `a'
at large flux levels ($\ge 10$ pfu and $\ge 100$ pfu).
But, such difference between them are very small.
For the SEP events with proton energy $\ge$ 50 MeV (Figure \ref{a}(b)),
the occurrence probabilities of SEP events in group `A' are all smaller than them
in group `a'. The difference between group `a' and `A' for the SEP events with proton
energy $\ge$ 50 MeV
are bigger than them  for the SEP events with proton energy $\ge$ 10 MeV
and became larger with the increasing of the flux level.
Even so, such difference are still small and less than 1$\sigma$.
Thus, the areas of corresponding CHs
did not show any evident influence on the CME in generating SEPs.

\begin{figure*}[tb]
\begin{center}
\includegraphics[width=0.8\hsize]{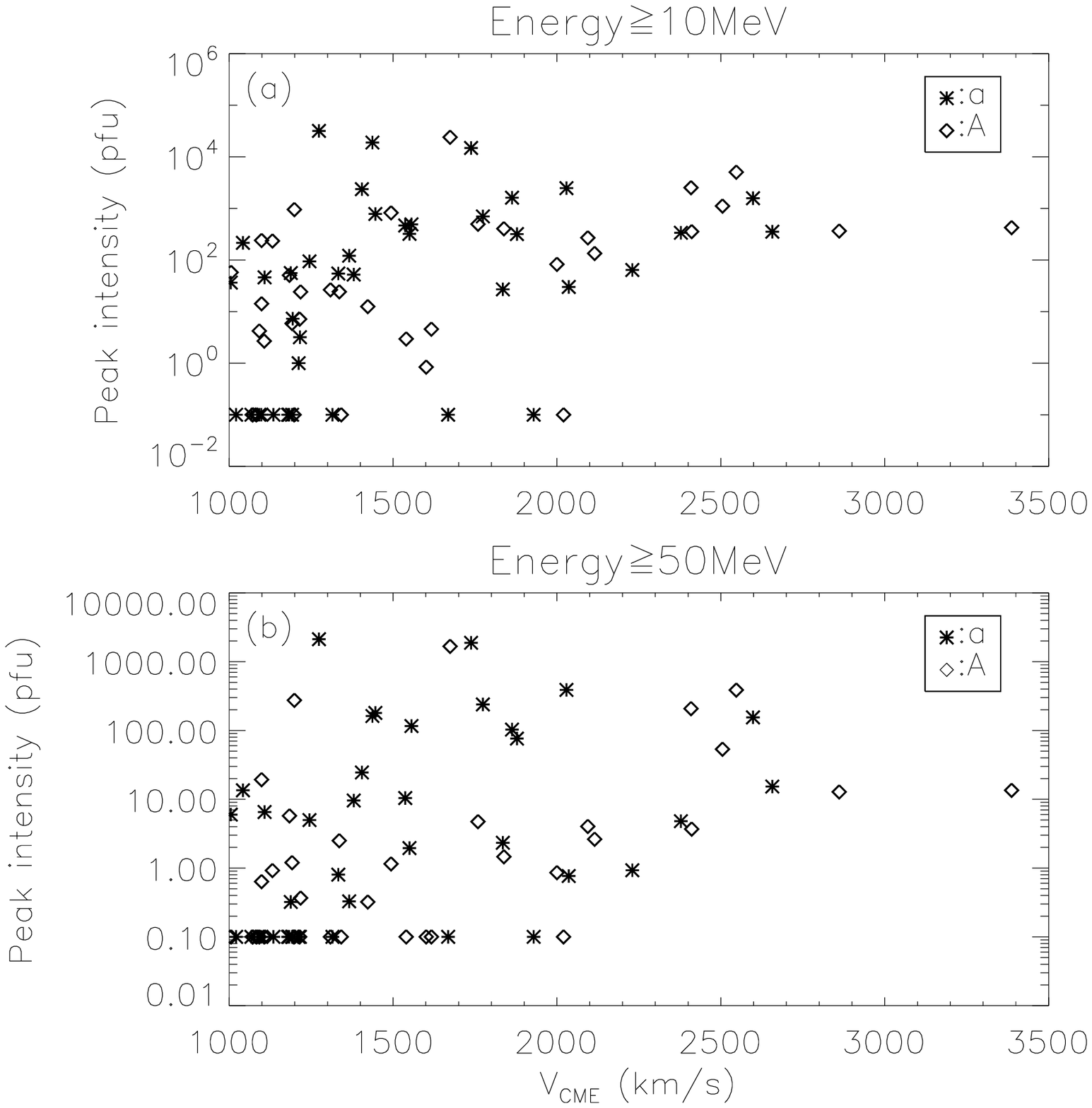}
\caption{The peak intensity of proton with energy $\ge$ 10MeV vs. associated CME speed.
The asterisks show the CMEs in group `a' while diamonds show the CME in group `A'.}
\label{a_v}
\end{center}
\end{figure*}

The peak intensity varied with the associated CME speed for group `a' and `A' are shown in
Figure \ref{a_v} while the mean value of the speed of SEPYCMEs and SEPNCMEs
are also listed in Table \ref{tb_cmev} (5th and 6th column). Similar with the analysis of CH proximity,
no obvious difference of CME speed distribution between group `a' and `A' could be found.
The mean value of the speed of SEPYCMEs and SEPNCMEs in these two groups are also similar.
This result confirms that the area of corresponding CHs show no evident
influence on CME in producing SEP event.

\subsection{The dependence of relative position}
\begin{figure*}[tb]
\begin{center}
\includegraphics[width=0.8\hsize]{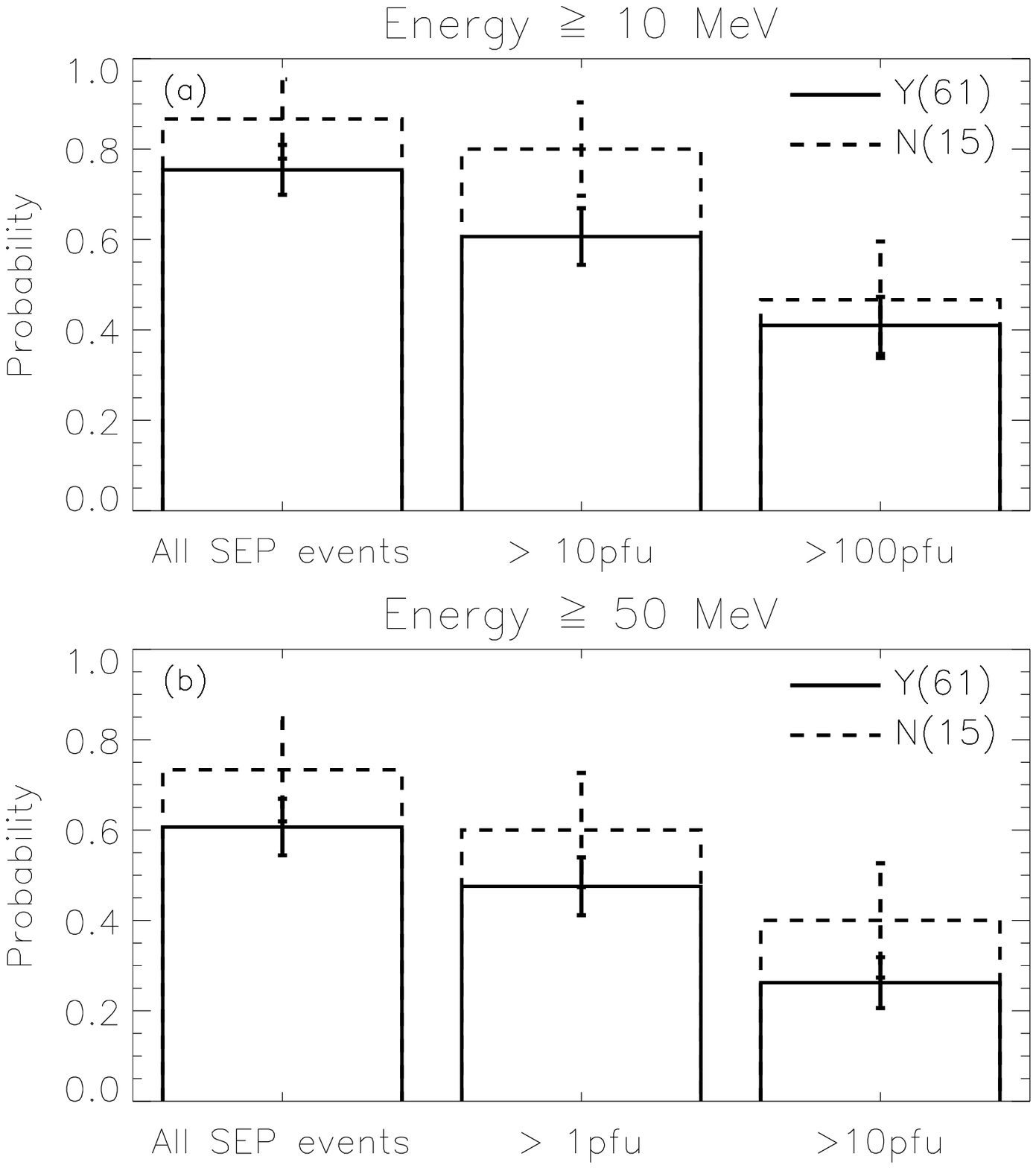}
\caption{Occurrence probabilities, $P$, of SEP events in terms of
relative position between CHs and CMEs for proton energies
$\ge$10 MeV and $\ge$50 MeV, respectively, } \label{y}
\end{center}
\end{figure*}
\begin{figure*}[tb]
\begin{center}
\includegraphics[width=0.8\hsize]{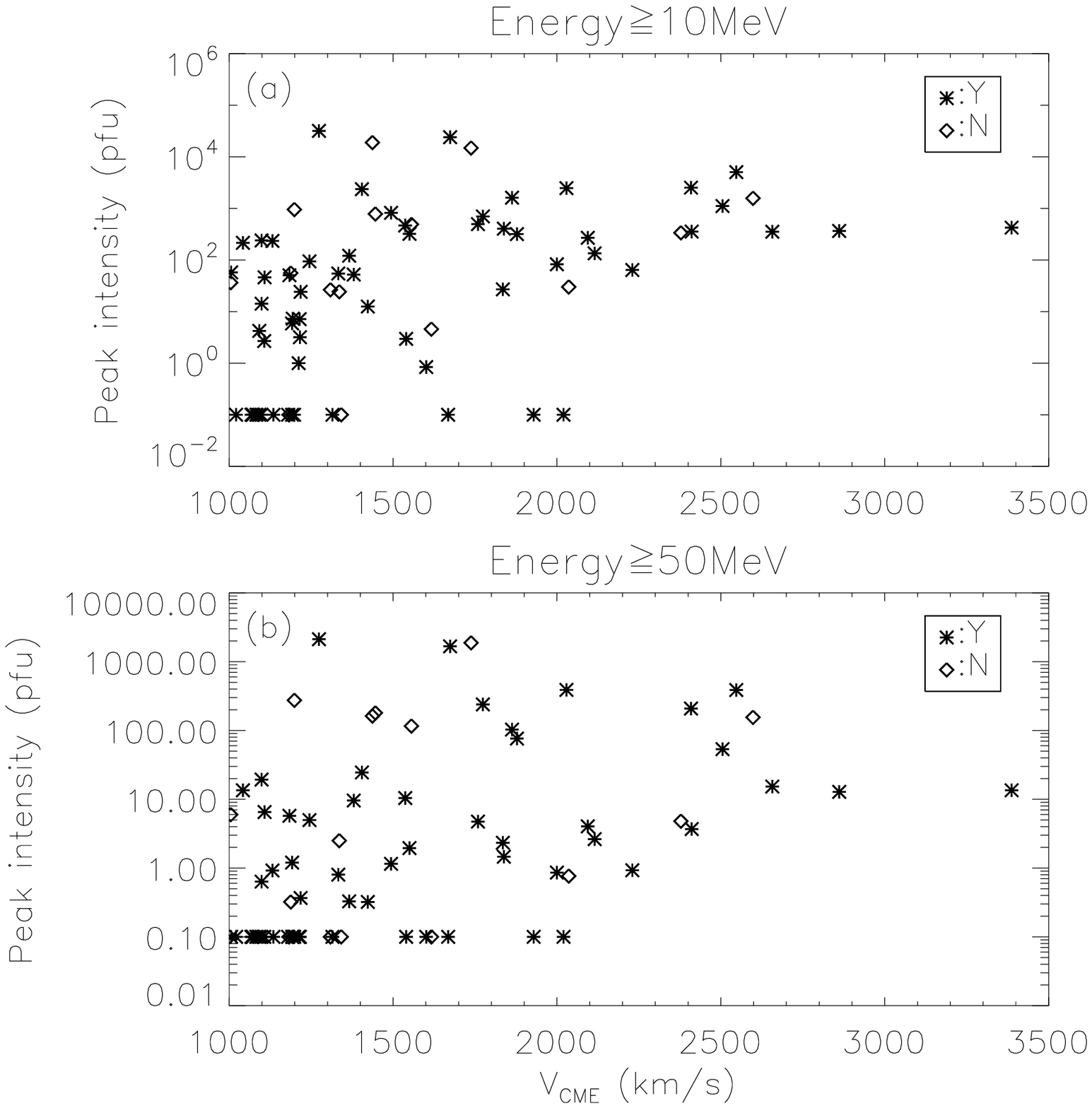}
\caption{The peak intensity of proton with energy $\ge$ 10MeV vs. associated CME speed.
The asterisks show the CMEs in group `Y' while diamonds show the CME in group `N'.
} \label{y_v}
\end{center}
\end{figure*}

Further, the possible impact of CHs location relative to the
corresponding CMEs is studied.
Figure \ref{y} shows the SEP occurrence probability of CMEs at different flux levels and different energy levels.
It is found that the SEP occurrence probability of CMEs at all flux levels  and energy levels
in group `Y' are smaller than them in group `N'. Specially, for the SEP events with flux level $\ge$10pfu
with proton energy $\ge$ 10MeV, the SEP occurrence in group `Y' is much small than it in group `N'.
The difference between these two groups is larger than 1$\sigma$ at this level.
But, such difference between these two groups are small and less than the
value of 1$\sigma$ for all the other levels.
The comparison of the speed of SEPYCMEs for group `Y' and `N'
is shown in Figure \ref{y_v}. As similar as we gotten in the analysis of CH proximity and CH area,
no obvious difference of the speed of SEPYCMEs between group `Y' and `N' could be found.
The average speed of SEPYCMEs is similar as the average speed of SEPNCMEs
as listed in last two columns of Table \ref{tb_cmev}.
These results imply that the relative location of CHs to the corresponding
CMEs has no evident effect on SEP events as the same as we get in Paper I.

\section{Summary and conclusions}
In order to study the influence of CHs on CMEs in producing SEP
events,  a total of 76 west-side fast halo CMEs during 1997 - 2008
are investigated, as well as their associated CHs. Different from
the CHs obtained by brightness method based on EIT 284\AA$ $ data in
paper I, the CHs we investigated in this paper are obtained with the
aid of the extrapolation of coronal magnetic field by CSSS model, in
which the MDI daily-updated synoptic magnetic field charts are
adopted as the bottom boundary condition. By using this method, all
the CHs, defined as the regions consisting of open magnetic field
lines only, over the entire solar surface are inferred.

After analyzing three parameters, CH proximity, area of corresponding CHs
and relative position between CHs and CMEs,
it is found that all of the statistical results do NOT have
significance exceeding the 1$\sigma$ level.
These parameters do NOT show any evident influence on SEP occurrence probability, and
the speed of SEPYCMEs also do NOT show any difference between different groups binarized by
these parameters.  These results confirmed the conclusion we got in Paper I and
\citet{Kahler_2004} that no evident influence of CHs on CME
in producing SEP events.

An expanding CME may drive a quasi-parallel shock at its flank as discussed by \citet{Kahler_2004}.
The condition of CME in driven shock in this situation is $V_{cme}$ larger
than local alf\'{v}en speed $V_a$ or sound speed $C_s$ only. Thus, the fast flow speed 
near CHs may show no influence on producing strong shock.
Beside, not only the plasma density but also the magnetic field strength in fast solar wind region is
smaller than them in slow solar wind region\citep{Ebert_etal_2009}, so the alf\'{v}en speed
 in fast solar wind region may not obvious faster than it in slow solar wind region.
Based on the these analysis, it could be expected that shock can also be produced in fast solar wind region near CH 
and no evident fast of the CME needed.
In addition, the shock interact with background solar wind may generate a turbulence. 
Such turbulence could be treated as the main mechanism that makes particles back to shock 
acceleration process to produce SEP events\citep{Reames_1999}. 
The close magnetic topology could only provide an addition method to make the particle back to shock 
acceleration\citep{Shen_etal_2008}. 
So, the influence of open magnetic field topology may weak in shock producing SEP events.

\normalem
\begin{acknowledgements}
We acknowledge the use of the data from the SOHO,
Yohkoh and GOES spacecraft, the CH maps from the Kitt Peak
Observatory. SOHO is a project of international cooperation between
ESA and NASA. This work is supported by grants from the National
Natural Science Foundation of China (40904046,40874075,40525014), the 973
National Basic Research Program (2006CB806304), Ministry of
Education of China (200530), the Program for NewCentury Excellent
Talents in University (NCET-08-0524) and the Chinese Academy of
Sciences (KZCX2-YW-QN511, KJCX2-YW-N28 and the startup fund). 
\end{acknowledgements}

\end{document}